\documentclass[12pt]{article}

\usepackage{amsmath}
\usepackage{cite}

\title{Lagrangian formulation of the massive higher spin
supermultiplets in three dimensional space-time}

\author{I.L. Buchbinder${}^{ab}$\thanks{joseph@tspu.edu.ru},
T.V. Snegirev${}^{ac}$\thanks{snegirev@tspu.edu.ru}, Yu.M.
Zinoviev$^d$\thanks{Yurii.Zinoviev@ihep.ru}
\\[0.5cm]
\it ${}^a$Department of Theoretical Physics,\\
\it Tomsk State Pedagogical University,\\
\it Tomsk 634061, Russia\\[0.3cm]
\it ${}^b$National Research Tomsk State University, Russia\\[0.3cm]
\it ${}^c$National Research Tomsk Polytechnical University,\\
Tomsk 634050, Russia\\[0.3cm]
\it ${}^d$Institute for High Energy Physics,\\
\it of National Research Center "Kurchatov Institute" \\
\it Protvino, Moscow Region, 142280, Russia}

\date{}

\begin{document}

\maketitle

\begin{abstract}
We give explicit construction for massive higher spin
supermultiplets for the case of minimal supersymmetry in $d=3$ and
find the corresponding Lagrangian formulations. We show that all
such massive supermultiplets can be straightforwardly constructed
out of the appropriately chosen set of massless ones exactly in the
same way as the gauge invariant description for the massive bosonic
(fermionic) field with spin $s$ can be obtained using a set of
massless fields with spins $s,s-1,\dots,0(1/2)$. Moreover, such
construction for the massive supermultiplets turns out to be perfectly
consistent with our previous results on the gauge invariant Lagrangian
formulation for massive higher spin bosons and fermions in $d=3$.
\end{abstract}

\thispagestyle{empty}
\newpage
\setcounter{page}{1}

\newpage
\section*{Introduction}

Last time the higher spin field theory is becoming one the the
central directions in modern theoretical and mathematical physics
(see e.g. the reviews 
\cite{Vas04,Sor04,BCIV05,FT08,BBS10,Sag11,DS14,Vas14}). In this paper
we are going to study some aspects of the higher spin field theory
related to constructing the massive supersymmetric higher spin models
in three space-time dimensions.\footnote{Strictly speaking, the notion
of spin does not exist in three dimensions in literal sense. For the
massless fields the only characteristic is statistics (i.e. boson or
fermion, see e.g. \cite{DJ91}), while massive fields are characterized
by their helicities in the same way as massless ones in four
dimensions. But the term spin is widely used in the literature on
three dimensional theories and so we will also use it here.}

In higher spin field theories, in spite of their infinite
dimensional gauge algebras, the supersymmetry still plays a
distinguished role in many aspects. It is enough to remind that, for
example, in the superstring theory all these massive higher spin
states are perfectly combined into the massive supermultiplets and
it is the supersymmetry that stays behind many nice feature of these
theories.

In general, the classification of massless and massive supermultiplets
is a rather straightforward algebraic task depending mainly on the
space-time dimension and on the specific properties of the fermions in
this space-time. But as far as explicit construction (in terms of
fields and Lagrangians) is concerned, the situation with massless and
massive higher spin supermultiplets drastically differ. For the
massless supermultiplets it is not hard to find such realization
following to the simple general pattern:
$$
\delta B \sim F \zeta, \qquad \delta F \sim \partial B \zeta.
$$
This is an essential reason why the supersymmetric massless higher
spin theory is developed good enough (see 
\cite{Cur79,KSP93,KS93,KS94,GKS96,GKS96a} for massless
higher spin models in four dimensions).

But the analogous construction for the massive supermultiplets
appears to be very complicated even if one uses a powerful
superfield technique (see e.g. 
\cite{BGLP02,BGLP02a,BGKP05,GK05,GKTM06,GK14} for some examples of
higher superspin superfields in four dimensional theory and
\cite{KTM11,KLRSTM13,KN14} in three dimensional ones). The reason is
that shifting from massless to the massive case one has to introduce
very complicated higher derivative corrections to the
supertransformations. Moreover the higher the spins of the fields
entering supermultiplet the higher the number of derivatives one has
to consider.

In four dimensions the solution for this problem was proposed
\cite{Zin02,Zin07,Zin07a} in component approach\footnote{The
Lagrangian formulation given in these papers is on-shell and does
not include the auxiliary fields which are needed for closing the
superalgebra. Off-shell supersymmetric formulation for these models
is still unknown.} based on the gauge invariant formalism for the
massive higher spin bosonic \cite{Zin01} and fermionic \cite{Met06}
fields. Recall that to obtain gauge invariant description for
massive bosonic (fermionic) field with spin $s$, one introduces a
set of massless fields with spins $s,s-1, \dots, 0(\frac{1}{2})$ with
their usual kinetic terms and local gauge transformations. Then one
adds all possible low derivative terms into the Lagrangian mixing
all these massless fields together as well as non-derivative
corrections to the gauge transformations to restore the gauge
symmetries broken by the mass terms. Now if one takes some massive
supermultiplet and decomposes each massive field into the
appropriate set of massless ones, one immediately sees that all
these massless fields are perfectly combined into the set of massless
supermultiplets. Thus the main idea of \cite{Zin02,Zin07,Zin07a} was
to generalize the gauge invariant description of massive particles
to the case of massive supermultiplets. Namely, one introduces
appropriate set of massless supermultiplets with all their fields,
kinetic terms and initial supertransformations and than adds lower
derivative terms to the Lagrangian as well {\bf non-derivative
corrections to the supertransformations for the fermions only}. It is
the absence of any higher derivative terms that makes such
construction to be pretty straightforward in spite of the large number
of fields involved. At the same time if one tries to fix all these
local symmetries then all these complicated higher derivatives
corrections reappeared just as the transformations restoring the
gauge. 

In this paper we give an explicit construction for the massive
supermultiplets with arbitrary spins in three dimensions for the case
of minimal supersymmetry. Naturally, this construction is based on the
gauge invariant description for the massive $d=3$ bosonic
\cite{BSZ12a} and fermionic \cite{BSZ14a} higher spin fields developed
in our previous works\footnote{Non gauge invariant formalism has been
developed in \cite{TV97}.}. Recall that in $d=4$ it was crucial for
the whole construction that there exists the possibility to consider a
kind of dual mixing for the massless supermultiplets:
$$
\left( \begin{array}{c} \Psi_{s+\frac{1}{2}} \\ A_s \end{array}
\right) \oplus \left( \begin{array}{c} B_s \\ \Phi_{s-\frac{1}{2}}
\end{array} \right) \quad \Rightarrow \quad \left( \begin{array}{ccc}
 & \Psi_{s+\frac{1}{2}} & \\ A_s & & B_s \\  & \Phi_{s-\frac{1}{2}} &
\end{array} \right),
$$
where $A_s$ and $B_s$ are two bosonic fields with equal spins but
opposite parities. Similarly, in $d=3$ case we found that it is
crucial that there exist the possibility to consider massless
supermultiplets containing one bosonic and two fermionic fields:
$$
\left( \begin{array}{c} \Psi_{s+\frac{1}{2}} \\ f_s \\
\Phi_{s-\frac{1}{2}} \end{array} \right).
$$
Recall that in three dimensions massless higher spin bosonic and
fermionic fields do not have any physical degrees of freedom. Thus
such a structure of the supermultiplet does not contradict to the fact
that in any supermultiplet the numbers of bosonic and fermionic
physical degrees of freedom must be equal. Note here that massive
higher spin supermultiplets do have physical degrees of freedom that
originate form the massless supermultiplets containing spins 1, 1/2
and 0 (that inevitably appear in the decomposition of massive
supermultiplet into the massless ones) exactly in the same way as in
the gauge invariant description of massive higher spin fields physical
degrees of freedom come from the components with spins 1, 1/2 and 0.

The paper is organized as follows. In Sections 1 and 2 we provide all
necessary formulas for massless high spin bosonic and fermionic fields
and massless higher spin supermultiplets in the frame-like multispinor
formalism we use in this work (see below). Section 3 contains two
relatively simple examples, namely massive supermultiplets
$(\frac{3}{2},1,\frac{1}{2})$ and
$(2,\frac{3}{2},\frac{3}{2})$\footnote{Such supermultiplet has been
constructed previously by dimensional reduction from $d=4$
\cite{BKPRYZ13}.}, illustrating our general technique. Sections 4 and
5
contains our main results: massive supermultiplets
$(s+\frac{1}{2},s,s-\frac{1}{2})$ and
$(s,s-\frac{1}{2},s-\frac{1}{2})$, correspondingly. To make our paper
self-contained as much as possible, we include three Appendices giving
gauge invariant description of massive higher spin bosons, fermions
with Majorana mass terms and fermions with Dirac mass terms, adopted
to the formalism used in this work.

\noindent
{\bf Notations and conventions} We will work in the frame-like
multispinor formalism where all objects are one-forms or zero-forms
completely symmetric on their local spinor indices. Spinor indices
$\alpha,\beta,\dots = 1,2$ are raised and lowered with the help of
antisymmetric by-spinor $\varepsilon^{\alpha\beta}$:
$$
\varepsilon^{\alpha\gamma} \varepsilon_{\gamma\beta} = -
\delta^\alpha{}_\beta, \qquad \varepsilon^{\alpha\beta} A_\beta =
A^\alpha, \qquad \varepsilon_{\alpha\beta} A^\beta = - A_\alpha.
$$
To simplify formulas we will often use a shorthand notations for
spinor indices:
$$
\Psi^{\alpha_1\alpha_2\dots \alpha_n} = \Psi^{\alpha(n)}.
$$
We will also assume that spinor indices denoted by the same letter and
placed on the same level are symmetrized:
$$
\Psi^{\alpha(n)} \zeta^\alpha = \Psi^{(\alpha_1\dots \alpha_n}
\zeta^{\alpha_{n+1})},
$$
where symmetrization contains the minimum number of terms necessary
without any normalization. Basis elements of $1,2,3$-form spaces are
respectively $e^{\alpha(2)}$, $E_2{}^{\alpha(2)}$, $E_3$ where the
last two are defined as double and triple wedge product of
$e^{\alpha(2)}$:
$$
e^{\alpha\alpha} \wedge e^{\beta\beta} =
\varepsilon^{\alpha\beta}{E}_2{}^{\alpha\beta},
$$
$$
E_2{}^{\alpha\alpha} \wedge e^{\beta\beta} = \varepsilon^{\alpha\beta}
\varepsilon^{\alpha\beta} E_3.
$$
Let us write some useful relations for these basis elements
$$
E_2{}^\alpha{}_\gamma \wedge e^{\gamma\beta} = 3
\varepsilon^{\alpha\beta} E_3, \qquad
e^\alpha{}_\gamma \wedge e^{\gamma\beta} = 4 E_2{}^{\alpha\beta}.
$$
In what follows we will systematically omit the $\wedge$ symbol.

\section{Massless fields}

In this section we give all necessary formulas for massless bosonic
and
fermionic higher spin fields in the flat three-dimensional
space-time.\\
{\bf Boson with spin $s=l+1$, $l \ge 1$} requires two one-forms
$\Omega^{\alpha(2l)}$ and $f^{\alpha(2l)}$ completely symmetric on
their local spinor indices. The free Lagrangian (which is a three-form
in our formalism) looks like:
\begin{equation}
(-1)^{l+1} {\cal L}_0 = l \Omega_{\alpha(2l-1)\beta} e^\beta{}_\gamma
\Omega^{\alpha(2l-1)\gamma} + \Omega_{\alpha(2l)} d
\Phi^{\alpha(2l)},
\end{equation}
where $d$ is an external derivative. This Lagrangian is invariant
under the following local gauge transformations:
\begin{equation}
\delta_0 \Omega^{\alpha(2l)} = d \eta^{\alpha(2l)}, \qquad
\delta_0 \Phi^{\alpha(2l)} = d \xi^{\alpha(2l)} + e^\alpha{}_\beta
\eta^{\alpha(2l-1)\beta},
\end{equation}
where $\eta^{\alpha(2l)}$ and $\xi^{\alpha(2l)}$ are zero forms also
completely symmetric in their indices. \\
{\bf Boson with spin 1} requires zero-form $B^{\alpha\beta}$ and
one-form $A$. The free Lagrangian and gauge transformations are:
\begin{equation}
{\cal L}_0 = E_3 B^{\alpha\beta} B_{\alpha\beta} - B_{\alpha\beta}
e^{\alpha\beta} d A, \qquad \delta_0 A = d \xi.
\end{equation}
Equation for the auxiliary field $B^{\alpha\beta}$ has the form:
\begin{equation}
2 E_3 B^{\alpha\beta} = e^{\alpha\beta} d A \quad \Rightarrow
\quad E_2{}^{\alpha\beta} B_{\alpha\beta} = d A
\end{equation}
and as a result we have the following on-shell identity:
$$
E^{\alpha\beta} d B_{\alpha\beta} = 0 \quad \Rightarrow \quad
E^\alpha{}_\gamma d B^{\beta\gamma} = E^\beta{}_\gamma d
B^{\alpha\gamma}.
$$
{\bf Boson with spin 0} requires two zero-forms $\pi^{\alpha\beta}$
and $\varphi$. The free Lagrangian:
\begin{equation}
{\cal L}_0 = E_3 \pi^{\alpha\beta} \pi_{\alpha\beta} -
E_2{}^{\alpha\beta} \pi_{\alpha\beta} d \varphi.
\end{equation}
Equation for the auxiliary field $\pi^{\alpha\beta}$:
\begin{equation}
2 E_3 \pi^{\alpha\beta} = E_2{}^{\alpha\beta} d \varphi \quad
\Rightarrow \quad e^{\alpha\beta} \pi_{\alpha\beta} = d \varphi
\end{equation}
leads to the following on-shell identity:
$$
e^{\alpha\beta} d \pi_{\alpha\beta} = 0 \quad \Rightarrow \quad
e^\alpha{}_\gamma d \pi^{\beta\gamma} = e^\beta{}_\gamma d
\pi^{\alpha\gamma}.
$$
{\bf Fermion with spin $s=l+\frac{1}{2}$, $l \ge 1$} is described by
one-form $\Psi^{\alpha(2l-1)}$ (also completely symmetric on spinor
indices) with the free Lagrangian and gauge transformations having the
form:
\begin{equation}
(-1)^{l+1} {\cal L}_0 = \frac{i}{2} \Psi_{\alpha(2l-1)} d
\Psi^{\alpha(2l-1)}, \qquad
\delta_0 \Psi^{\alpha(2l-1)} = d \zeta^{\alpha(2l-1)}.
\end{equation}
{\bf Fermion with spin $\frac{1}{2}$} is described by zero-form
$\phi^\alpha$ with the free Lagrangian:
\begin{equation}
{\cal L}_0 = \frac{i}{2} \phi_\alpha E^\alpha{}_\beta d \phi^\beta.
\end{equation}

\section{Massless supermultiplets}

As it has been already noted in the Introduction, the main idea of
this work is that massive supermultiplet can be straightforwardly
constructed out of appropriate set of massless super\-multiplets
exactly in the same way as gauge invariant description of massive
bosonic or fermionic higher spin field can be constructed out of
appropriate set of massless ones. As it will be seen further on in the
analysis of massive supermultiplets, the main role as a building block
is played by the massless supermultiplet containing one bosonic and
two fermionic fields, namely (here and in what follows $s$ is always
an integer) $(s+\frac{1}{2},s,s-\frac{1}{2})$. \\
{\bf Supermultiplet $(l+\frac{3}{2},l+1,l+\frac{1}{2})$, $l \ge 1$}
contains two fermionic $\Phi^{\alpha(2l+1)}$ and $\Psi^{\alpha(2l-1)}$
and two bosonic $\Omega^{\alpha(2l)}$, $f^{\alpha(2l)}$ one-forms. The
sum of their kinetic terms:
\begin{eqnarray}
{\cal L}_0 &=& (-1)^{l+1} [ l \Omega_{\alpha(2l-1)\beta}
e^\beta{}_\gamma \Omega^{\alpha(2l-1)\gamma} + \Omega_{\alpha(2l)} d
f^{\alpha(2l)} \nonumber \\
 && \qquad + \frac{i}{2} \Phi_{\alpha(2l+1)} d \Phi^{\alpha(2l+1)}
- \frac{i}{2} \Psi_{\alpha(2l-1)} d \Psi^{\alpha(2l-1)} ]
\end{eqnarray}
is invariant under the following global supertransformations:
\begin{eqnarray}
\delta f^{\alpha(2l)} &=& i\alpha_l \Psi^{\alpha(2l-1)} \zeta^\alpha +
i(2l+1) \beta_l \Phi^{\alpha(2l)\beta} \zeta_\beta, \nonumber \\
\delta \Phi^{\alpha(2l+1)} &=& \beta_l \Omega^{\alpha(2l)}
\zeta^\alpha, \\
\delta \Psi^{\alpha(2l-1)} &=& 2l\alpha_l \Omega^{\alpha(2l-1)\beta}
\zeta_\beta. \nonumber
\end{eqnarray}
In what follows we will fix the normalization of supertransformations
so that
$$
2l\alpha_l{}^2 + (2l+1) \beta_l{}^2 = 2,
$$
while the relative values of $\alpha_l$ and $\beta_l$ will depend on
the massive supermultiplet which this massless one enters in. Thus in
general such supermultiplet, its Lagrangian and the
supertransformations contain two fermionic and one bosonic fields. But
as the particular cases we can obtain supermultiplets with one
fermionic and one bosonic fields. Namely, putting $\alpha_l = 0$ we
get supermultiplet ($l+3/2,l+1$), while the case $\beta_l = 0$
corresponds to ($l+1,l+1/2$).\\
\\
{\bf Supermultiplet $(\frac{3}{2},1,\frac{1}{2})$} contains fermionic
one-form $\Phi^\alpha$ and zero-form $\psi^\alpha$ as well as bosonic
zero-form $B^{\alpha\beta}$ and one-form $A$. The sum of their kinetic
terms
\begin{equation}
{\cal L}_0 = - \frac{i}{2} \Phi_\alpha d \Phi^\alpha + E
B^{\alpha\beta} B_{\alpha\beta} - B_{\alpha\beta} e^{\alpha\beta} d A
+ \frac{i}{2} \psi_\alpha E^\alpha{}_\beta d \psi^\alpha
\end{equation}
is invariant under the following
supertransformations\footnote{Strictly speaking this Lagrangian is
invariant up to the terms proportional to the auxiliary field
$B^{\alpha\beta}$ equation only. Thus there are two possible approach
here. From one hand one can introduce non-trivial corrections to the
supertransformations for this auxiliary field. Another possibility,
that we will systematically follow here and further on, is to use
equations for the auxiliary fields in calculating all variations.}:
\begin{eqnarray}
\delta A &=& i\beta_0 \Phi_\alpha \zeta^\alpha + i\alpha_0 \psi_\alpha
e^{\alpha\beta} \zeta_\beta, \nonumber \\
\delta \Phi^\alpha &=& - \beta_0 e^{\beta\gamma} B_{\beta\gamma}
\zeta^\alpha, \\
\delta \psi^\alpha &=& 4\alpha_0 B^{\alpha\beta} \zeta_\beta.
\nonumber
\end{eqnarray}
{\bf Supermultiplet $(\frac{1}{2},0)$} contains fermionic zero-form
$\phi_\alpha$ and two bosonic zero-forms $\pi^{\alpha\beta}$ and
$\varphi$. The sum of kinetic terms
\begin{equation}
{\cal L}_0 = \frac{i}{2} \phi_\alpha E^\alpha{}_\beta d \phi^\alpha -
E \pi^{\alpha\beta} \pi_{\alpha\beta} + E^{\alpha\beta}
\pi_{\alpha\beta} d \varphi
\end{equation}
is invariant (provided one takes into account the equation for the
auxiliary field $\pi^{\alpha\beta}$) under the following
supertransformations:
\begin{equation}
\delta \varphi = i\tilde{\beta}_0 \phi_\alpha \zeta^\alpha, \qquad
\delta \phi^\alpha = 2\tilde{\beta}_0 \pi^{\alpha\beta} \zeta_\beta.
\end{equation}

\section{Simple examples}

\subsection{Massive supermultiplet $(\frac{3}{2},1,\frac{1}{2})$}

Gauge invariant description of massive spin-$\frac{3}{2}$ requires
massless spin-$\frac{3}{2}$ and spin-$\frac{1}{2}$ ones, while massive
spin-1 requires massless spin-1 and spin-0. Thus to construct massive
supermultiplet $(\frac{3}{2},1,\frac{1}{2})$ we need two massless
supermultiplets $(\frac{3}{2},1,\frac{1}{2})$ and $(\frac{1}{2},0)$:
$$
\left( \begin{array}{c} \frac{3}{2} \\ 1 \\ \frac{1}{2} \end{array}
\right) \quad \Rightarrow \quad \left( \begin{array}{c} \frac{3}{2} \\
1 \\ \frac{1}{2} \end{array} \right) \oplus \left( \begin{array}{c}
\frac{1}{2} \\ 0 \end{array} \right).
$$
We begin with the sum of kinetic terms for all necessary fields:
\begin{eqnarray}
{\cal L}_0 &=& - \frac{i}{2} \Phi_\alpha d \Phi^\alpha + E
B^{\alpha\beta} B_{\alpha\beta} - B_{\alpha\beta} e^{\alpha\beta} d A
+ \frac{i}{2} \psi_\alpha E^\alpha{}_\beta d \psi^\beta \nonumber \\
 && + \frac{i}{2} \phi_\alpha E^\alpha{}_\beta d \phi^\beta - E
\pi^{\alpha\beta} \pi_{\alpha\beta} + E^{\alpha\beta}
\pi_{\alpha\beta} d \varphi
\end{eqnarray}
as well as their initial supertransformations:
\begin{eqnarray}
\delta_0 A &=& i\beta_0 \Phi_\alpha \zeta^\alpha + i\alpha_0
\psi_\alpha e^{\alpha\beta} \zeta_\beta, \nonumber \\
\delta_0 \Phi^\alpha &=& - \beta_0 e^{\beta\gamma} B_{\beta\gamma}
\zeta^\alpha, \qquad \delta_0 \psi^\alpha = 4\alpha_0 B^{\alpha\beta}
\zeta_\beta, \\
\delta_0 \varphi &=& i\tilde{\beta}_0 \phi_\alpha \zeta^\alpha, \qquad
\delta_0 \phi^\alpha = 2\tilde{\beta}_0 \pi^{\alpha\beta} \zeta_\beta.
\nonumber
\end{eqnarray}
Now we add the most general low derivative terms:
\begin{eqnarray}
{\cal L}_1 &=& 2m E_{\alpha\beta} \pi^{\alpha\beta} A + ia_1
\Phi_\alpha e^\alpha{}_\beta \Phi^\beta + ia_2 \Phi_\alpha
E^\alpha{}_\beta \psi^\beta + ia_3 \Phi_\alpha E^\alpha{}_\beta
\phi^\beta \nonumber \\
 && + ia_4 E \psi_\alpha \psi^\alpha + ia_5 E \psi_\alpha
\phi^\alpha + ia_6 E \phi_\alpha \phi^\alpha.
\end{eqnarray}
This breaks the invariance under the supertransformations producing
\begin{eqnarray*}
\delta_0 {\cal L}_1 &=& - 2i(2a_1\beta_0-a_2\alpha_0) \Psi_\alpha
E^{(\alpha}{}_\beta B^{\gamma)\beta} \zeta_\gamma + 2ia_2\alpha_0
\Psi_\alpha (EB) \zeta^\alpha \\
 && + ia_3\tilde{\beta}_0 \Psi_\alpha E^{(\alpha}{}_\beta
\pi^{\gamma)\beta} \zeta_\gamma + i(2m\beta_0+a_3\tilde{\beta}_0)
\Psi_\alpha (E\pi) \zeta^\alpha \\
 && - 2i(a_2\beta_0-4a_4\alpha_0) E \psi_\alpha B^{\alpha\beta}
\zeta_\beta + 2i(2m\alpha_0+a_5\tilde{\beta}_0) E \psi_\alpha
\pi^{\alpha\beta} \zeta_\beta \\
 && - 2i(a_3\beta_0-2a_5\alpha_0) E \phi_\alpha B^{\alpha\beta}
\zeta_\beta + im\tilde{\beta}_0 \phi_\alpha e^{\alpha\beta} d A
\zeta_\beta + 4ia_6\tilde{\beta}_0 E \phi_\alpha \pi^{\alpha\beta}
\zeta_\beta.
\end{eqnarray*}
Recall that, as we have already mentioned above, all calculations are
performed up to the terms proportional to the auxiliary fields
equations and in this case one has take into account that equation for
the $\pi^{\alpha\beta}$ field was modified and looks like:
\begin{equation}
E^\alpha{}_\beta \pi^{\alpha\beta} = - \frac{1}{2} e^{\alpha(2)} [ d
\varphi + 2m A ].
\end{equation}
To cancel these new variations we put
$$
2a_1\beta_0 = a_2\alpha_0, \qquad 2m\beta_0 = - a_3\tilde{\beta}_0,
\qquad a_2\beta_0 = 4a_4\alpha_0, \qquad 2a_3\beta_0 -
4a_5\alpha_0 = 2m\tilde{\beta}_0
$$
and introduce the following corrections to the
supertransformations\footnote{Let us stress that here and in what
follows the only corrections we have to introduce in our approach are
the non-derivative ones to the fermionic fields.}:
\begin{equation}
\delta_1 \Psi^\alpha = \gamma_1 A \zeta^\alpha + \gamma_2 \varphi
e^{\alpha\beta} \zeta_\beta, \qquad \delta_1 \psi_\alpha = \gamma_3
\varphi \zeta_\alpha, \qquad \delta_1 \phi_\alpha = \gamma_4 \varphi
\zeta_\alpha.
\end{equation}
Then all variations with one derivative vanish provided
$$
\gamma_1 = 2a_2\alpha_0, \qquad
\gamma_2 = \frac{a_3\tilde{\beta}_0}{2}, \qquad
\gamma_3 = 2m\alpha_0 + a_5\tilde{\beta}_0, \qquad
\gamma_4 = 2a_6\tilde{\beta}_0
$$
leaving us with variations without derivatives:
\begin{eqnarray*}
\delta_1 {\cal L}_1 &=& i(2a_1\gamma_1 + ma_3\alpha_1)
\Psi_\alpha e^\alpha{}_\beta A \zeta^\beta +
i(8a_1\gamma_2 - a_2\gamma_3 - a_3\gamma_4) \Psi_\alpha
E^{\alpha\beta} \varphi \zeta_\beta \\
 && + i(a_2\gamma_1 + 2m\gamma_3) \psi_\alpha E^{\alpha\beta} A
\zeta_\beta - i(3a_2\gamma_2 - 2a_4\gamma_3 - a_5\gamma_4) E
\psi_\alpha \varphi \zeta^\alpha \\
 && + i(a_3\gamma_1 + 2m\gamma_4) \phi_\alpha E^{\alpha\beta} A
\zeta_\beta - i(3a_3\gamma_2 - a_5\gamma_3 - 2a_6\gamma_4) E
\phi_\alpha \varphi \zeta^\alpha.
\end{eqnarray*}
Thus we obtain:
$$
a_1 = a_4 = a_6 = \frac{m}{2}, \qquad a_2 = - a_3 = \sqrt{2}m, \qquad
 a_5 = - 2m,
$$
$$
\tilde{\beta}_0{}^2 = 2\beta_0{}^2 = 4 \alpha_0{}^2.
$$
Now if we introduce new variables
$$
\tilde{\psi}^\alpha = \frac{1}{\sqrt{2}}(\psi^\alpha - \phi^\alpha),
\qquad \tilde{\phi}^\alpha = \frac{1}{\sqrt{2}} ( \psi^\alpha +
\phi^\alpha),
$$
then the fermionic mass terms take the form
\begin{equation}
{\cal L}_m = \frac{im}{2} [ \Phi_\alpha e^\alpha{}_\beta \Phi^\alpha +
4 \Phi_\alpha E^\alpha{}_\beta \tilde{\psi}^\beta + 3 E
\tilde{\psi}_\alpha \tilde{\psi}^\alpha ] - \frac{im}{2} E
\tilde{\phi}_\alpha \tilde{\phi}^\alpha,
\end{equation}
which corresponds to massive spin-$\frac{3}{2}$ (in the gauge
invariant formalism with the field $\tilde{\psi}^\alpha$ playing the
role of Stueckelberg one) and massive spin-$\frac{1}{2}$ with equal
masses. Note that though we begin with the most general form of the
fermionic mass terms, the supersymmetry leads us to their form
corresponding to gauge invariant formulation of massive
spin-$\frac{3}{2}$. In what follows from the very beginning we will
use gauge invariant description of massive bosonic and fermionic
higher spin fields entering supermultiplets. It will greatly simplify
all calculations and always happens to be compatible with the
supersymmetry.

\subsection{Massive supermultiplet $(2,\frac{3}{2},\frac{3}{2})$}

This supermultiplet has been constructed previously by dimensional
reduction from four dimensions \cite{BKPRYZ13}. Here we will show how
its construction can be worked out in our approach. In the massless
limit massive spin-2 decomposes into massless spin-2, spin-1 and
spin-0 ones, while massive spin-$\frac{3}{2}$ into massless
spin-$\frac{3}{2}$ and spin-$\frac{1}{2}$. Thus in this case the
decomposition of massive supermultiplet into the massless ones has the
form:
$$
\left( \begin{array}{ccc} & 2 & \\ \frac{3}{2} && \frac{3}{2}
\end{array} \right) \quad \Rightarrow \quad \left( \begin{array}{c} 2
\\ \frac{3}{2} \end{array} \right) \oplus \left( \begin{array}{c}
\frac{3}{2} \\ 1 \\ \frac{1}{2} \end{array} \right) \oplus \left(
\begin{array}{c} \frac{1}{2} \\ 0 \end{array} \right).
$$
Again we begin with the sum of kinetic terms for all necessary
fields:
\begin{eqnarray}
{\cal L}_0 &=& \Omega_{\alpha\beta} e^\beta{}_\gamma
\Omega^{\alpha\gamma} + \Omega_{\alpha\beta} d f^{\alpha\beta} -
\frac{i}{2} \Psi_\alpha d \Psi^\alpha \nonumber \\
 && - \frac{i}{2} \Phi_\alpha d \Phi^\alpha + E
B^{\alpha\beta} B_{\alpha\beta} - B_{\alpha\beta} e^{\alpha\beta} d A
+ \frac{i}{2} \psi_\alpha E^\alpha{}_\beta d \psi^\beta \nonumber \\
 && + \frac{i}{2} \phi_\alpha E^\alpha{}_\beta d \phi^\beta - E
\pi^{\alpha\beta} \pi_{\alpha\beta} + E^{\alpha\beta}
\pi_{\alpha\beta} d \varphi
\end{eqnarray}
as well as their initial supertransformations:
\begin{eqnarray}
\delta f^{\alpha\beta} &=& i\alpha_1 \Psi^{(\alpha} \zeta^{\beta)},
\qquad \delta \Psi^\alpha = 2\alpha_1 \Omega^{\alpha\beta}
\zeta_\beta, \nonumber \\
\delta A &=& i\beta_0 \Phi_\alpha \zeta^\alpha + i\alpha_0
\psi_\alpha e^{\alpha\beta} \zeta_\beta, \\
\delta \Phi^\alpha &=& - \beta_0 e^{\beta\gamma} B_{\beta\gamma}
\zeta^\alpha, \qquad \delta \psi^\alpha = 4\alpha_0 B^{\alpha\beta}
\zeta_\beta, \nonumber \\
\delta \varphi &=& i\tilde{\beta}_0 \phi_\alpha \zeta^\alpha, \qquad
\delta \phi^\alpha = 2\tilde{\beta}_0 \pi^{\alpha\beta} \zeta_\beta.
\nonumber
\end{eqnarray}
Now we have to add low derivative terms. As we have already mentioned,
we will use gauge invariant description both for the massive spin-2 as
well as massive spin-$\frac{3}{2}$ fields. As for the massive spin-2
here the choice is unambiguous --- we have just one spin-1 and spin-0
zero fields to the roles of Stueckelberg ones. And for the two massive
spin-$\frac{3}{2}$ fields by analogy with four-dimensional case
\cite{Zin02} we will assume that two Majorana fields will combine into
a Dirac one. Thus we introduce the following terms (see Appendices A
and C):
\begin{eqnarray}
{\cal L}_1 &=& - 2m e_{\alpha\beta} \Omega^{\alpha\beta} A - m
f_{\alpha\beta} E^\beta{}_\gamma B^{\alpha\gamma} - 4m
E_{\alpha\beta} \pi^{\alpha\beta} A \nonumber \\
 && + im [ \Psi_\alpha e^\alpha{}_\beta \Phi^\beta + 2
\Psi_\alpha E^\alpha{}_\beta \psi^\beta + 2 \Phi_\alpha
E^\alpha{}_\beta \phi^\beta + 3 E \psi_\alpha \phi^\alpha ], \\
{\cal L}_2 &=& \frac{m^2}{4} f_{\alpha\beta} e^\beta{}_\gamma
f^{\alpha\gamma} - m^2 E^{\alpha\beta} f_{\alpha\beta} \varphi +
\frac{3m^2}{2} E \varphi^2.
\end{eqnarray}
As in the previous case calculating all variations one has to use
equations for the auxiliary fields which in this case have the form:
\begin{eqnarray}
0 &=& e^\alpha{}_\beta \Omega^{\alpha\beta} + d f^{\alpha(2)} + 2m
e^{\alpha(2)} A, \nonumber \\
0 &=& 2 E B^{\alpha(2)} - e^{\alpha(2)} d A + \frac{m}{2}
E^\alpha{}_\beta f^{\alpha\beta}, \\
0 &=& 2E \pi^{\alpha(2)} - E^{\alpha(2)} [ d \varphi - 4m A ].
\nonumber
\end{eqnarray}
To compensate for variations with one derivative $\delta_0 {\cal L}_1$
we introduce the following corrections to the supertransformations:
\begin{equation}
\delta_1 \Psi^\alpha = \gamma_1 A \zeta^\alpha, \qquad
\delta_1 \Phi^\alpha = \gamma_2 f^{\alpha\beta} \zeta_\beta + \gamma_3
\varphi e^{\alpha\beta} \gamma_\beta, \qquad
\delta_1 \psi^\alpha = \gamma_4 \varphi \zeta^\alpha,
\end{equation}
where
$$
\gamma_1 = 4m\alpha_0, \qquad
\gamma_2 = - m\alpha_1, \qquad
\gamma_3 = m\tilde{\beta}_0, \qquad
\gamma_4 = 3m\tilde{\beta}_0 + 4m\alpha_0.
$$
Then all variations without derivatives $\delta_1 {\cal L}_1 +
\delta_0 {\cal L}_2$ vanish provided
$$
\alpha_1{}^2 = \tilde{\beta}_0{}^2 = 4\alpha_0{}^2, \qquad
\beta_0{}^2 = \alpha_0{}^2.
$$

\section{Massive supermultiplet $(s+\frac{1}{2},s,s-\frac{1}{2})$}

In this case the same line of reasoning leads us to the decomposition:
$$
\left( \begin{array}{c} s+\frac{1}{2} \\ s \\ s-\frac{1}{2}
\end{array} \right) \quad \Rightarrow \quad \sum_{l=1}^{s}
\left( \begin{array}{c} l + \frac{1}{2} \\ l \\ l-\frac{1}{2}
\end{array} \right) \oplus \left( \begin{array}{c} \frac{1}{2} \\ 0
\end{array} \right).
$$
Correspondingly we begin with appropriate sum of kinetic terms for all
fields
\begin{eqnarray}
{\cal L}_0 &=& \sum_{l=1}^{s-1} (-1)^{l+1} [ l
\Omega_{\alpha(2l-1)\beta} e^\beta{}_\gamma
\Omega^{\alpha(2l-1)\gamma} + \Omega_{\alpha(2l)} d f^{\alpha(2l)} ]
\nonumber \\
 && + E B_{\alpha(2)} B^{\alpha(2)} - B_{\alpha(2)} e^{\alpha(2)} d A
- E \pi_{\alpha(2)} \pi^{\alpha(2)} + \pi_{\alpha(2)} E^{\alpha(2)} d
\varphi \nonumber \\
 && + \frac{i}{2} [ \sum_{l=0}^{s-2} (-1)^{l+1}
\Psi_{\alpha(2l+1)} d \Psi^{\alpha(2l+1)} + \sum_{l=0}^{s-1}
(-1)^{l+1} \Phi_{\alpha(2l+1)} d \Phi^{\alpha(2l+1)} ] \nonumber \\
 && + \frac{i}{2} \phi_\alpha E^\alpha{}_\beta d \phi^\beta +
\frac{i}{2} \chi_\alpha E^\alpha{}_\beta d \chi^\beta \label{lag0}
\end{eqnarray}
and their initial supertransformations:
\begin{eqnarray}
\delta f^{\alpha(2l)} &=& i\alpha_l \Psi^{\alpha(2l-1)} \zeta^\alpha +
i(2l+1)\beta_l \Phi^{\alpha(2l)\beta} \zeta_\beta, \qquad l \ge 1
\nonumber \\
\delta_0 \Phi^{\alpha(2l+1)} &=& \beta_l \Omega^{\alpha(2l)}
\zeta^\alpha, \qquad
\delta_0 \Psi^{\alpha(2l-1)} = 2l\alpha_l \Omega^{\alpha(2l-1)\beta}
\zeta_\beta, \nonumber \\
\delta A &=& i\beta_0 \Phi_\alpha \zeta^\alpha + i\alpha_0 \psi_\alpha
e^{\alpha\beta} \zeta_\beta, \label{var0} \\
\delta \Phi^\alpha &=& - \beta_0 e^{\beta(2)} B_{\beta(2)}
\zeta^\alpha, \qquad \delta \psi^\alpha = 4\alpha_0 B^{\alpha\beta}
\zeta_\beta, \nonumber \\
\delta \varphi &=& i\tilde{\beta}_0 \phi_\alpha \zeta^\alpha, \qquad
\delta \phi^\alpha = 2\tilde{\beta}_0 \pi^{\alpha\beta} \zeta_\beta.
\nonumber
\end{eqnarray}
Now we have to add the lower derivative terms. For the bosonic terms
we take the ones corresponding to gauge invariant description of
massive spin-$s$ boson (see Appendix A), while for the fermionic terms
we introduce the most general ones compatible with the fact that they
have to correspond to gauge invariant description of two massive
fermions with spin-$(s+\frac{1}{2})$ and spin-$(s-\frac{1}{2})$ with
equal masses (see Appendix B):
\begin{eqnarray}
{\cal L}_{1b} &=& \sum_{l=1}^{s-2} (-1)^{l+1} a_l [ - \frac{(l+2)}{l}
\Omega_{\alpha(2l)\beta(2)} e^{\beta(2)} f^{\alpha(2l)} +
\Omega_{\alpha(2l)} e_{\beta(2)} f^{\alpha(2l)\beta(2)} ] \nonumber \\
 && + a_0 [ 2 \Omega_{\alpha(2)} e^{\alpha(2)} A - f_{\alpha\beta}
E^\beta{}_\gamma B^{\alpha\gamma} ] + \tilde{a}_0 \pi_{\alpha(2)}
E^{\alpha(2)} A, \label{lag1} \\
{\cal L}_2 &=& \sum_{l=1}^{s-1} (-1)^{l+1} c_l
f_{\alpha(2l-1)\beta} e^\beta{}_\gamma f^{\alpha(2l-1)\gamma} +
\tilde{c}_1 h_{\alpha(2)} E^{\alpha(2)} \varphi + c_0 E \varphi^2,
\label{lag2}
\end{eqnarray}
\begin{eqnarray}
{\cal L}_{1f} &=& i \sum_{l=0}^{s-1} (-1)^{l+1} \tilde{b}_l
\Phi_{\alpha(2l)\beta} e^\beta{}_\gamma \Phi^{\alpha(2l)\gamma} + i
\sum_{l=0}^{s-2} (-1)^{l+1} [ b_l \Phi_{\alpha(2l)\beta}
e^\beta{}_\gamma \Psi^{\alpha(2l)\gamma} + \hat{b}_l
\Psi_{\alpha(2l)\beta}  e^\beta{}_\gamma  \Psi^{\alpha(2l)\gamma} ]
\nonumber \\
 && + iE [ \tilde{b}_{-1} \phi_\alpha \phi^\alpha + b_{-1} \phi_\alpha
\psi^\alpha + \hat{b}_{-1} \psi_\alpha \psi^\alpha ] \nonumber \\
 && + i \sum_{l=1}^{s-1} (-1)^{l+1} \Phi_{\alpha(2l-1)\beta(2)}
e^{\beta(2)} [ \tilde{d}_l \Phi^{\alpha(2l-1)} + e_l
\Psi^{\alpha(2l-1)} ] \nonumber \\
 && + i \sum_{l=1}^{s-2} (-1)^{l+1} \Psi_{\alpha(2l-1)\beta(2)}
e^{\beta(2)} [ d_l \Psi^{\alpha(2l-1)} + \tilde{e}_l
\Phi^{\alpha(2l-1)} ] \nonumber \\
 && + i \Phi_\alpha E^\alpha{}_\beta [ \tilde{d}_0 \phi^\beta +
e_0 \psi^\beta ] + i \Psi_\alpha E^\alpha{}_\beta [ d_0
\psi^\beta + \tilde{e}_0 \phi^\beta ]. \label{lagf}
\end{eqnarray}
As in the previous cases, calculating the variations we use auxiliary
field equations:
$$
0 = e^\alpha{}_\beta \Omega^{\alpha(2l-1)\beta} + d f^{\alpha(2l)} +
a_l e_{\beta(2)} f^{\alpha(2l)\beta(2)} +
\frac{(l+1)a_{l-1}}{l(l-1)(2l-1)} e^{\alpha(2)} f^{\alpha(2l-2)}.
$$
From the variations with one derivative we found that we must
introduce the full set of corrections to the supertransformations:
\begin{eqnarray}
\delta_1 \Phi^{\alpha(2l+1)} &=& \tilde{\gamma}_l
f^{\alpha(2l+1)\beta} \zeta_\beta + \tilde{\delta}_l f^{\alpha(2l)}
\zeta^\alpha, \nonumber \\
\delta_1 \Psi^{\alpha(2l+1)} &=& \gamma_l f^{\alpha(2l+1)\beta}
\zeta_\beta + \delta_l f^{\alpha(2l)} \zeta^\alpha, \nonumber \\
\delta_1 \Phi^\alpha &=& \tilde{\gamma}_0 f^{\alpha\beta} \zeta_\beta
+ \tilde{\delta}_0 A \zeta^\alpha
+ \rho_0 e^{\alpha\beta} \varphi \zeta_\beta, \\
\delta_1 \Psi^\alpha &=& \delta_0 f^{\alpha\beta} \zeta_\beta +
\gamma_0 A \zeta^\alpha + \tilde{\tilde{\rho}}_0 e^{\alpha\beta}
\varphi \zeta_\beta ,\nonumber \\
\delta_1 \psi^\alpha &=& \tilde{\rho}_0 \varphi \zeta^\alpha, \qquad
\delta_1 \phi^\alpha = \hat{\rho}_0 \varphi \zeta^\alpha. \nonumber
\end{eqnarray}
Moreover, supersymmetry requires that we put $\tilde{e}_l = 0$ and it
is this constraint that allowed us to find the solution for the
fermionic mass terms in Appendix B. Now when all the coefficients in
the Lagrangian are fixed, it is straightforward to find solution for
the parameters of the supertransformations:
\begin{eqnarray}
\alpha_l{}^2 &=& \frac{(l+1)}{l(2l+1)}, \qquad
\beta_l{}^2 = \frac{2l}{(2l+1)^2}, \\
\alpha_0{}^2 &=& \frac{1}{4}, \qquad
\beta_0{}^2 = \frac{1}{2}, \qquad
\tilde{\beta}_0{}^2 = 1. \nonumber
\end{eqnarray}
where we set the normalization so that
$$
2l \alpha_l{}^2 + (2l+1) \beta_l{}^2 = 2.
$$
Then the parameters determining corrections to the
supertransformations are also fixed:
\begin{eqnarray}
\gamma_l{}^2 &=& \frac{(s+l+1)(s-l-1)}{2l(l+1)^2(2l+1)^2} m^2,
\nonumber \\
\tilde{\gamma}_l{}^2 &=& \frac{(s+l+1)(s-l-1)}{(l+1)(l+2)(2l+3)} m^2,
\nonumber \\
\delta_l{}^2 &=& \frac{(l+1)}{(l+2)(2l+3)} m^2, \\
\tilde{\delta}_l{}^2 &=& \frac{m^2}{2l(2l+1)^2}, \nonumber
\end{eqnarray}
$$
\gamma_0{}^2 = 2(s-1)(s+1)m^2, \qquad
\tilde{\delta}_0{}^2 = 2m^2,
$$
$$
\rho_0{}^2 = \tilde{\rho}_0{}^2 = s^2m^2, \qquad
\hat{\rho}_0{}^2 = m^2, \qquad
\tilde{\tilde{\rho}}_0 = 0.
$$

\section{Massive supermultiplet $(s,s-\frac{1}{2},s-\frac{1}{2})$}

In this case the decomposition looks like:
$$
\left( \begin{array}{c} s \\ s-\frac{1}{2}, s-\frac{1}{2}
\end{array} \right) \quad \Rightarrow \quad \left( \begin{array}{c} s
\\ s-\frac{1}{2} \end{array} \right) \oplus
\sum_{l=1}^{s-1} \left( \begin{array}{c} l + \frac{1}{2} \\ l \\
l-\frac{1}{2} \end{array} \right) \oplus \left( \begin{array}{c}
\frac{1}{2} \\ 0 \end{array} \right).
$$
Thus we need the same set of fields as in the previous case except the
field $\Phi^{\alpha(2s-1)}$ (recall that the supermultiplet
($s,s-1/2$) is just a particular case of ($s+1/2,s,s-1/2$) one). So we
take the same massless Lagrangian (\ref{lag0}) with this field omitted
and the same set of initial supertransformations (\ref{var0}) where
now $\beta_{s-1}=0$. As far as the low derivative terms, the bosonic
terms will again have the same form (\ref{lag1}) and (\ref{lag2}),
while by analogy with four dimensional case \cite{Zin07a} we will
assume that fermions have Dirac mass terms compatible with gauge
invariant description (see Appendix C):
\begin{eqnarray}
{\cal L}_{1f} &=& i \sum_{l=0}^{s-2} (-1)^{l+1} b_l
\Psi_{\alpha(2l)\beta} e^\beta{}_\gamma \Phi^{\alpha(2l)\gamma} +
i \tilde{b}_0 E \phi_\alpha \chi^\alpha \nonumber \\
 && + i \sum_{l=1}^{s-2} (-1)^{l+1} [ d_l \Psi_{\alpha(2l-1)\beta(2)}
e^{\beta(2)} \Psi^{\alpha(2l-1)} + \tilde{d}_l
\Phi_{\alpha(2l-1)\beta(2)} e^{\beta(2)} \Phi^{\alpha(2l-1)} ]
\nonumber \\
 && + id_0 \Psi_\alpha E^\alpha{}_\beta \psi^\beta + i\tilde{d}_0
\Phi_\alpha E^\alpha{}_\beta \phi^\beta. \label{lagf2}
\end{eqnarray}
Calculating all variations with one derivative $\delta_0 {\cal L}_1$
we
find that we have to introduce the following corrections to
supertransformations:
\begin{eqnarray}
\delta_1 \Psi^{\alpha(2l+1)} &=& \gamma_l f^{\alpha(2l)} \zeta^\alpha,
\qquad \delta \Phi^{\alpha(2l+1)} = \tilde{\gamma}_l
f^{\alpha(2l+1)\beta} \zeta_\beta, \nonumber \\
\delta_1 \Psi^\alpha &=& \gamma_0 A \zeta^\alpha, \qquad
\delta_1 \Phi^\alpha = \tilde{\gamma}_0 f^{\alpha\beta} \zeta_\beta +
\rho_0 e^{\alpha\beta} \varphi \zeta_\beta, \\
\delta_1 \psi^\alpha &=& \tilde{\rho}_0 \varphi \zeta^\alpha \nonumber
\end{eqnarray}
and obtain the following expressions for the parameters determining
supertransformations:
\begin{eqnarray}
\alpha_l{}^2 &=& \frac{(l+1)(s+l)}{2sl(2l+1)}, \qquad
\beta_l{}^2 = \frac{l(s-l-1)}{s(2l+1)^2}, \nonumber \\
\alpha_0{}^2 &=& \frac{1}{8}, \qquad
\beta_0{}^2 = \frac{(s-1)}{4s}, \qquad
\tilde{\beta}_0{}^2 = \frac{1}{2},
\end{eqnarray}

\begin{eqnarray}
\gamma_l{}^2 &=& \frac{s(s-l-1)m^2}{4l(l+1)^2(2l+1)^2}, \nonumber \\
\tilde{\gamma}_l{}^2 &=& \frac{s(s+l+1)m^2}{2(l+1)(l+2)(2l+3)},
\nonumber \\
\gamma_0{}^2 &=& s(s-1)m^2, \qquad
\tilde{\gamma}_0{}^2 = \frac{s(s+1)m^2}{12}, \\
\rho_0{}^2 &=& \frac{s(s-1)m^2}{4}, \qquad
\tilde{\rho}_0{}^2 = 4(s-1)^2m^2. \nonumber
\end{eqnarray}

\section*{Conclusion}

In this paper we have constructed the minimal supersymmetric
Lagrangian formulation for all massive supermultiplets with
arbitrary spins in $d=3$. We have shown that as in the $d=4$ case
such massive supermultiplets can be straightforwardly built out 
of the appropriately chosen set of massless ones. Such
procedure can be considered as a supersymmetric generalization for
the gauge invariant formalism for massive higher spin bosonic and
fermionic fields where the description for the massive field is
obtained through the set of the massless ones. In most cases
constructing the Lagrangians we from the very beginning choose mass
terms compatible with such gauge invariant description for massive
fields. But as we have shown in one case and checked in others even
if one starts with the most general form of the mass terms without
any preliminary assumptions the supersymmetry alone will unavoidably
lead to such form. Thus the very idea of gauge invariant description
for the massive higher spin fields is in the perfect agreement with
the supersymmetry.

As the directions of the further development one can point out: {\bf
1.} The approach can be applied to the extended supersymmetries as
well. {\bf 2.} It would be interesting to consider interaction of
such massive higher spin supermultiplets with supergravity. {\bf 3.}
The approach considered in this paper is on-shell. It would be
interesting to develop a completely off-shell superfield Lagrangian
formulation for the three-dimensional supersymmetric massive higher
spin theories. Some preliminary results have already been obtained
in \cite{KUZ1}\footnote{Recently we have been informed by S.M.
Kuzenko that he has unpublished yet results on superfield
formulation for massless three-dimensional $N=2$ supersymmetric models
with arbitrary superspins.}. {\bf 4.} We have constructed the
supersymmetric massive higher spin models in flat 3d space It would be
interesting to generalize the models under consideration to $AdS_3$
space and apply it to problem of $AdS_3/CFT_2$ duality.

\section*{Acknowledgments}
The authors thank S.Deser, S.M. Kuzenko and M.A. Vasiliev for useful
comments. I.L.B and T.V.S are grateful to the grant for LRSS,
project No. 88.2014.2 and RFBR grant, project No. 15-02-03594-a for
partial support. Their research was also supported by Russian
Ministry of Education and Science, project 2014/387.122. T.V.S
acknowledges partial support from the President of Russia grant for
young scientists No. MK-6453.2015.2 and RFBR grant No. 14-02-31254.
Work of Yu.M.Z was supported in parts by RFBR grant No. 14-02-01172.

\appendix

\section{Massive boson with spin $s \ge 2$}

Gauge invariant description of massive boson with spin $s$
\cite{BSZ12a} requires massless fields with spins $s \ge l \ge 0$.
Thus we introduce a collection of one-forms $f^{\alpha(2l)}$,
$\Omega^{\alpha(2l)}$, $s-1 \ge l \ge 1$ as well as one-form $A$ and
zero-forms $B^{\alpha\beta}$, $\pi^{\alpha\beta}$ and $\varphi$. As
usual, we begin with the sum of kinetic terms for all fields:
\begin{eqnarray}
{\cal L}_0 &=& \sum_{l=1}^{s-1} (-1)^{l+1} [ l
\Omega_{\alpha(2l-1)\beta} e^\beta{}_\gamma
\Omega^{\alpha(2l-1)\gamma} + \Omega_{\alpha(2l)} d f^{\alpha(2l)}
] \nonumber \\
 && + E B_{\alpha\beta} B^{\alpha\beta} - B_{\alpha\beta}
e^{\alpha\beta} d A - E \pi_{\alpha\beta} \pi^{\alpha\beta} +
\pi_{\alpha\beta} E^{\alpha\beta} d \varphi
\end{eqnarray}
and their initial gauge transformations:
\begin{equation}
\delta_0 f^{\alpha(2l)} = d \xi^{\alpha(2l)} + e^\alpha{}_\beta
\eta^{\alpha(2l-1)\beta}, \qquad \delta_0 \Omega^{\alpha(2l)} = d
\eta^{\alpha(2l)}, \qquad \delta_0 A = d \xi.
\end{equation}
To proceed we add the most general terms with one derivative:
\begin{eqnarray}
{\cal L}_1 &=&  \sum_{l=1}^{s-2} (-1)^{l+1} [ b(l)
\Omega_{\alpha(2)\beta(2l)} e^{\alpha(2)} f^{\beta(2l)} + a(l)
\Omega_{\alpha(2l)} e_{\beta(2)} f^{\alpha(2l)\beta(2)} ] \nonumber \\
 && - a_0 \Omega_{\alpha\beta} e^{\alpha\beta} A - b_0
f_{\alpha\beta} E^\beta{}_\gamma B^{\alpha\gamma} + \tilde{a}_0
\pi_{\alpha\beta} E^{\alpha\beta} A
\end{eqnarray}
as well as the most general corrections to the gauge transformations:
\begin{eqnarray}
\delta_1 \Omega^{\alpha(2l)} &=& \kappa_1(l) e_{\beta(2)}
\eta^{\alpha(2l)\beta(2)} + \kappa_2(l) e^{\alpha(2)}
\eta^{\alpha(2l-2)} + \kappa_3(l) e^\alpha{}_\beta
\xi^{\alpha(2l-1)\beta}, \nonumber \\
\delta_1 \Phi^{\alpha(2l)} &=& \kappa_4(l) e_{\beta(2)}
\xi^{\alpha(2l)\beta(2)} + \kappa_5(l) e^{\alpha(2)}
\xi^{\alpha(2l-2)}, \nonumber \\
\delta_1 B^{\alpha\beta} &=& \kappa_1(0) \eta^{\alpha\beta}, \qquad
\delta_1 A = D \xi + \kappa_4(0) e_{\alpha\beta} \xi^{\alpha\beta}, \\
\delta_1 \pi^{\alpha\beta} &=& \kappa_6 \xi^{\alpha\beta}, \qquad
\delta_1 \varphi = \kappa_7 \xi. \nonumber
\end{eqnarray}
All variations coming from $\delta_0 {\cal L}_1 + \delta_1 {\cal L}_0$
vanish provided
$$
\kappa_1(l) = - b(l), \qquad
\kappa_2(l) = \frac{a(l-1)}{l(2l-1)}, \qquad
\kappa_4(l) = a(l),
$$
$$
\kappa_5(l) = - \frac{b(l-1)}{l(2l-1)}, \qquad
b(l) = - \frac{(l+2)a(l)}{l},
$$
$$
\kappa_1(0) = - a_0, \qquad
\kappa_4(0) = \frac{b_0}{4}, \qquad
\kappa_7 = - \tilde{a}_0, \qquad
a_0 + 2b_0 = 0.
$$
To proceed we introduce the most general terms without derivatives:
\begin{equation}
{\cal L}_2 = \sum_{l=1}^{s-1} (-1)^{l+1} c(l) f_{\alpha(2l-1)\beta}
e^\beta{}_\gamma f^{\alpha(2l-1)\gamma} + \tilde{c}_1
f_{\alpha\beta} E^{\alpha\beta} \varphi + c_0 E \varphi^2.
\end{equation}
Then all remaining variations can be canceled if we put
$$
\frac{2(k+2)(2k+3)}{(k+1)(2k+1)} a(k)^2 -
\frac{2(k+1)}{(k-1)} a(k-1)^2 + 4c(k) = 0,
$$
$$
\kappa_3(l) = \frac{c(l)}{l}, \qquad
(l+2)^2 c(l+1) = l(l+1)c(l),
$$
$$
\kappa_6 = \tilde{c}_1 = - \frac{\tilde{a}_0b_0}{4}, \qquad
\tilde{a}_0{}^2 = 64c_1, \qquad
b_0{}^2 = \frac{(s+1)(s-1)}{3} m^2,
$$
where we choose normalization
$$
m^2 = \frac{2s(s-1)}{(s-2)} a(s-2)^2.
$$
The last relation in the second line is just a recurrent relation
on $c(l)$ and it gives
$$
c(l) = \frac{s^2(s-1)}{l(l+1)^2} c(s-1).
$$
Than the first line can be considered as a recurrent relation on
$a(l)$. For $l = s-1$ we obtain
$$
c(s-1) = \frac{m^2}{4(s-1)},
$$
and then we get general solution
$$
a(l)^2 = \frac{l(s+l+1)(s-l-1)}{2(l+1)(l+2)(2l+3)} m^2, \qquad
a_0{}^2 = \frac{(s-1)(s+1)}{3} m^2.
$$

\section{Massive fermions with Majorana mass terms}

To construct gauge invariant description of massive fermion with
spin-$(s+\frac{1}{2})$ \cite{BSZ14a} we introduce a set of one-forms
$\Psi^{\alpha(2l+1)}$, $s-1 \ge l \ge 0$ and zero-form $\chi^\alpha$.
As for the Lagrangian we take the sum of kinetic terms for all fields
as well as the most general form for the mass-like terms:
\begin{eqnarray}
{\cal L}_0 &=& \frac{i}{2} \sum_{l=0}^{s-1} (-1)^{l+1}
\Psi_{\alpha(2l+1)} d \Psi^{\alpha(2l+1)} + \frac{i}{2} \chi_\alpha
E^\alpha{}_\beta d \chi^\beta \nonumber \\
 && + i \sum_{l=0}^{s-1} (-1)^{l+1} b_l \Psi_{\alpha(2l)\beta}
e^\beta{}_\gamma \Psi^{\alpha(2l)\gamma} + i\tilde{b}_0 E \chi_\alpha
\chi^\alpha \nonumber \\
 && + i \sum_{l=1}^{s-1} (-1)^{l+1} d_l \Psi_{\alpha(2l-1)\beta(2)}
e^{\beta(2)} \Psi^{\alpha(2l-1)} + id_0 \Psi_\alpha E^\alpha{}_\beta
\chi^\beta.
\end{eqnarray}
At the same time we introduce the most general ansatz for the local
gauge transformations:
\begin{eqnarray}
\delta \Psi^{\alpha(2l+1)} &=& d \xi^{\alpha(2l+1)} + \alpha_l
e^\alpha{}_\beta \xi^{\alpha(2l)\beta} + \beta_l e^{\alpha(2)}
\xi^{\alpha(2l-1)} + \gamma_l e_{\beta(2)} \xi^{\alpha(2l+1)\beta(2)},
\nonumber \\
\delta \chi^\alpha &=& \tilde{\alpha}_0 \xi^\alpha.
\end{eqnarray}
For variations with one derivative to vanish we have to put
$$
\alpha_l = \frac{2b_l}{(2l+1)}, \qquad
\beta_l = \frac{d_l}{l(2l+1)}, \qquad
\gamma_l = d_{l+1}, \qquad
\tilde{\alpha}_0 = d_0.
$$
Than all variations without derivatives vanish provided
$$
\frac{4b_l{}^2}{(2l+1)} - d_l{}^2 + \frac{(l+2)(2l+1)}{(l+1)(2l+3)}
d_{l+1}{}^2 = 0, \qquad b_{l-1} = \frac{(2l+3)}{(2l+1)} b_l.
$$
These relations can be easily solved and give
$$
b_l = \frac{(2s+1)}{2(2l+3)} m, \qquad
\tilde{b}_0 = - 3b_0,
$$
$$
d_l{}^2 = \frac{(s-l)(s+l+1)}{2(l+1)(2l+1)} m^2, \qquad
d_0{}^2 = 2s(s+1)m^2.
$$
where we choose normalization by setting $b_{s-1} = \frac{m}{2}$. In
what follows we will need analogous solution for the fermion with
spin-$(s-\frac{1}{2})$ which has the form
$$
b_l = \frac{(2s-1)}{2(2l+3)} m, \qquad
\tilde{b}_0 = - 3b_0,
$$
$$
d_l{}^2 = \frac{(s-l-1)(s+l)}{2(l+1)(2l+1)} m^2, \qquad
d_0{}^2 = 2s(s-1)m^2.,
$$

For the massive supermultiplet ($s+\frac{1}{2},s,s-\frac{1}{2})$) we
need two massive fermions with spin-$(s+\frac{1}{2})$ and
$(s-\frac{1}{2})$. The simplest solution is to take just the sum of
corresponding mass terms:
\begin{eqnarray}
{\cal L}_1 &=& i \sum_{l=0}^{s-1} \frac{(2s+1)m}{2(2l+3)}
\tilde{\Phi}_{\alpha(2l)\beta} e^\beta{}_\gamma
\tilde{\Phi}^{\alpha(2l)\gamma} - i \sum_{l=0}^{s-2}
\frac{(2s-1)m}{2(2l+3)} \tilde{\Psi}_{\alpha(2l)\beta}
e^\beta{}_\gamma \tilde{\Psi}^{\alpha(2l)\gamma} \nonumber \\
 && + i \sum_{l=1}^{s-1} A(s,l) \tilde{\Phi}_{\alpha(2l-1)\beta(2)}
e^{\beta(2)} \tilde{\Phi}^{\alpha(2l-1)}
- i \sum_{l=1}^{s-2} A(s-1,l) \tilde{\Psi}_{\alpha(2l-1)\beta(2)}
e^{\beta(2)} \tilde{\Psi}^{\alpha(2l-1)} \nonumber \\
 && + im \sqrt{2s(s+1)} \tilde{\Phi}_\alpha E^\alpha{}_\beta
\tilde{\phi}^\beta - im\sqrt{2s(s-1)} \tilde{\Psi}_\alpha
E^\alpha{}_\beta \tilde{\psi}^\beta \nonumber \\
 && - \frac{i(2s+1)m}{2} E \tilde{\phi}_\alpha \tilde{\phi}^\alpha +
\frac{i(2s-1)m}{2} E \tilde{\psi}_\alpha \tilde{\psi}^\alpha,
\end{eqnarray},
where we denote
$$
A(s,l) = \sqrt{\frac{(s-l)(s+l+1)}{2(l+1)(2l+1)}} m.
$$
But in general the variables in terms of which the mass terms turn out
to be diagonal do not coincide with the ones entering massless
supermultiplets. The most general situation corresponds to the
possible mixings for the pairs of fermions with equal spins. Thus we
introduce:
\begin{eqnarray}
\tilde{\Phi}_{\alpha(2s-1)} &=& \Phi_{\alpha(2s-1)}, \nonumber \\
\tilde{\Phi}_{\alpha(2l+1)} &=& \cos \theta_l \Phi_{\alpha(2l+1)} +
\sin \theta_l \Psi_{\alpha(2l+1)}, \nonumber \\
\tilde{\Psi}_{\alpha(2l+1)} &=& - \sin \theta_l \Phi_{\alpha(2l+1)} +
\cos \theta_l \Psi_{\alpha(2l+1)}, \\
\tilde{\phi}^\alpha &=& \cos \theta \phi^\alpha + \sin \theta
\psi^\alpha, \nonumber \\
\tilde{\psi}^\alpha &=& - \sin \theta \phi^\alpha + \cos \theta
\psi^\alpha. \nonumber
\end{eqnarray}
Than for the mass terms we obtain the Lagrangian (\ref{lagf}) used in
our construction of corresponding massive supermultiplet, where
\begin{eqnarray*}
\tilde{b}_{s-1} &=& \frac{m}{2}, \\
\tilde{b}_l &=& \frac{(2s+1) \cos^2 \theta_l - (2s-1) \sin^2
\theta_l}{2(2l+3)} m, \\
b_l &=& \frac{4s \sin \theta_l \cos \theta_l }{(2l+3)} m \\
\hat{b}_l &=& \frac{(2s+1) \sin^2 \theta_l - (2s-1) \cos^2
\theta_l}{2(2l+3)} m, \\
\tilde{b}_{-1} &=& - \frac{(2s+1)\cos^2 \theta-(2s-1)\sin^2
\theta}{2}m, \\
b_{-1} &=& - 4s \sin \theta \cos \theta m \\
\hat{b}_{-1} &=& - \frac{(2s+1)\sin^2 \theta - (2s-1)\cos^2 \theta}{2}
m,
\end{eqnarray*}
\begin{eqnarray*}
\tilde{d}_{s-1} &=& \frac{m}{\sqrt{(2s-1)}} \cos \theta_{s-2}, \\
\tilde{d}_l &=& A(s,l) \cos \theta_l \cos \theta_{l-1} - A(s-1,l) \sin
\theta_l \sin \theta_{l-1}, \\
\tilde{d}_0 &=& \sqrt{2s(s+1)} m \cos \theta_0 \cos \theta -
\sqrt{2s(s-1)} m \sin \theta_0 \sin \theta, \\
d_l &=& A(s,l) \sin \theta_l \sin \theta_{l-1} - A(s-1,l) \cos
\theta_l \cos \theta_{l-1}, \\
d_0 &=& \sqrt{2s(s+1)} m \sin \theta_0 \sin \theta -
\sqrt{2s(s-1)} m \cos \theta_0 \cos \theta, \\
e_{s-1} &=& \frac{m}{\sqrt{(2s-1)}} \sin \theta_{s-2}, \\
e_l &=& A(s,l) \cos \theta_l \sin \theta_{l-1} + A(s-1,l) \sin
\theta_l \cos \theta_{l-1}, \\
e_0 &=& \sqrt{2s(s+1)} m \cos \theta_0 \sin \theta +
\sqrt{2s(s-1)} m \sin \theta_0 \cos \theta, \\
\tilde{e}_l &=& A(s,l) \sin \theta_l \cos \theta_{l-1} + A(s-1,l) \cos
\theta_l \sin \theta_{l-1}, \\
\tilde{e}_0 &=& \sqrt{2s(s+1)} m \sin \theta_0 \cos \theta +
\sqrt{2s(s-1)} m \cos \theta_0 \sin \theta.
\end{eqnarray*}
But supersymmetry requires that $\tilde{e}_l=0$ and this gives a
recurrent relation on the mixing angles:
$$
\tan \theta_{l-1} = - \sqrt{\frac{(s-l)(s+l+1)}{(s-l-1)(s+l)}}
\tan \theta_l.
$$
For the massive supermultiplet ($s+\frac{1}{2},s,s-\frac{1}{2})$) we
will need the following simple solution for this relation:
$$
\sin \theta_l = (-1)^l \sqrt{\frac{s-l-1}{2s}}, \qquad
\cos \theta_l = \sqrt{\frac{s+l+1}{2s}}.
$$
Than for the coefficients in the fermionic mass terms (\ref{lagf}) we
obtain:
\begin{eqnarray*}
\tilde{b}_l &=& - \tilde{b}_{-1} = - \hat{b}_{-1} = \frac{m}{2},
\qquad \hat{b}_l = - \frac{2l+1}{2(2l+3)} m, \\
b_l &=& (-1)^l \frac{2\sqrt{(s-l-1)(s+l+1)}}{(2l+3)} m, \qquad
b_{-1} = 2sm, \\
\tilde{d}_l &=& \sqrt{\frac{(s-l)(s+l)}{2(l+1)(2l+1)}} m, \qquad
\tilde{d}_0 = \sqrt{2} sm, \\
d_l &=& - \sqrt{\frac{(s-l-1)(s+l+1)}{2(l+1)(2l+1)}} m, \qquad
d_0 = - \sqrt{2(s+1)(s-1)} m, \\
e_l &=& - (-1)^l \sqrt{\frac{1}{2(l+1)(2l+1)}} m, \qquad
e_0 = - \sqrt{2} m.
\end{eqnarray*}

\section{Massive fermions with Dirac mass terms}

For the massive supermultiplet $(s,s-\frac{1}{2},s-\frac{1}{2})$ we
need a pair of massive fermions with spin-$(s-\frac{1}{2})$ with equal
masses and with mass-like terms having a Dirac form. The kinetic
terms has the usual form:
\begin{equation}
{\cal L}_0 = \frac{i}{2} \sum_{l=0}^{s-2} (-1)^{l+1} [
\Psi_{\alpha(2l+1)} d \Psi^{\alpha(2l+1)} + \Phi_{\alpha(2l+1)} d
\Phi^{\alpha(2l+1)} ] + \frac{i}{2} \psi_\alpha E^\alpha{}_\beta d
\psi^\beta + \frac{i}{2} \phi_\alpha E^\alpha{}_\beta d \phi^\beta,
\end{equation}
while for the mass-like terms we choose the Lagrangian (\ref{lagf2}).
For the field's $\Psi^{\alpha(2l+1)}$ local gauge transformations we
consider the following ansatz:
\begin{eqnarray*}
\delta \Psi^{\alpha(2l+1)} &=& d \xi^{\alpha(2l+1)} + \beta_l
e^{\alpha(2)} \xi^{\alpha(2l-1)} + \gamma_l e_{\beta(2)}
\xi^{\alpha(2l+1)\beta(2)}, \\
\delta \Phi^{\alpha(2l+1)} &=& \alpha_l e^\alpha{}_\beta
\xi^{\alpha(2l)\beta}, \qquad \delta \psi^\alpha = \tilde{\alpha}_0
\xi^\alpha.
\end{eqnarray*}
For variations with one derivative to cancel we have to put
$$
\alpha_l = \frac{b_l}{(2l+1)}, \qquad
\beta_l = \frac{d_l}{l(2l+1)}, \qquad
\gamma_l = d_{l+1}, \qquad
\tilde{\alpha}_0 = d_0.
$$
Then all variations without derivatives vanish provided
$$
\frac{b_l{}^2}{(2l+1)} - d_l{}^2 + \frac{(l+2)(2l+1)}{(l+1)(2l+3)}
d_{l+1}{}^2 = 0,
$$
$$
\frac{(2l+3)}{(2l+1)} b_l\tilde{d}_l -  b_{l-1}d_l = 0.
$$
Analogously, the invariance under the local gauge transformations for
the $\Phi^{\alpha(2l+1)}$ field gives:
$$
\frac{b_l{}^2}{(2l+1)} - \tilde{d}_l{}^2 +
\frac{(l+2)(2l+1)}{(l+1)(2l+3)} \tilde{d}_{l+1}{}^2 = 0,
$$
$$
\frac{(2l+3)}{(2l+1)} b_ld_l -  b_{l-1}\tilde{d}_l = 0.
$$
From these equations we obtain the following solution:
$$
b_l = \frac{(2s-1)}{(2l+3)} m, \qquad
d_l{}^2 = \tilde{d}_l{}^2 = \frac{(s-l-1)(s+l)}{2(l+1)(2l+1)} m^2,
$$
$$
\tilde{b}_0 = 3b_0, \qquad
d_0{}^2 = \tilde{d}_0{}^2 = 2s(s-1)m^2,
$$
where we set $b_{s-2} = m$.

\end{document}